# Sustainable Incentives for Mobile Crowdsensing
## (Auctions, Lotteries, and Trust and Reputation Systems)

Tie Luo [a], Salil S. Kanhere [b], Jianwei Huang [c], Sajal K. Das [d], and Fan Wu [e]

[a] Institute for Infocomm Research, A*STAR, Singapore
[b] School of Computer Science and Engineering, The University of New South Wales, Australia
[c] Department of Information Engineering, The Chinese University of Hong Kong, Hong Kong, China
[d] Department of Computer Science, Missouri University of Science and Technology, USA
[e] Department of Computer Science, Shanghai Jiao Tong University, China

*Abstract*—Proper incentive mechanisms are critical for mobile crowdsensing systems to motivate people to actively and persistently participate. This article provides an exposition of design principles of six incentive mechanisms, drawing special attention to the sustainability issue. We cover three primary classes of incentive mechanisms: auctions, lotteries, and trust and reputation systems, as well as three other frameworks of promising potential: bargaining games, contract theory, and market-driven mechanisms.

## I. INTRODUCTION

Mobile crowdsensing (MCS) is a new crowdsourcing technique that exploits the sensing capabilities of personal mobile devices, such as smartphones and wearables, to collect data from a large group of individuals. It is advantageous in low deployment cost and vast geographic coverage, and has found numerous applications in diverse domains including transportation, environment monitoring, smart city, and pervasive healthcare. However, MCS systems often face the challenge of *insufficient participation* due to two reasons: (i) sensing incurs nontrivial costs to participants in terms of battery consumption, mobile data usage, time, and effort, and (ii) sensor-data collection may not have direct benefit to participants but often requires long-term commitment. Therefore, designing proper *incentive mechanisms* is pivotal to motivate the crowd to participate in and sustain MCS.

This tutorial article provides an exposition of six incentive mechanisms (Fig. 1) that can be applied to MCS. This area of study is fascinating due to its interdisciplinary nature: auctions and lotteries are deeply rooted in microeconomics, while trust and reputation systems are a subject of artificial intelligence by tradition; bargaining games, contract theory, and market-driven mechanisms all sit on the boundary between economics and computer science. This article elaborates the first three mechanisms in length due to their wide adoption in the literature, but we also summarize the salient technical features of the other three because of their promising potential.

Compared to its predecessor, *crowdsourcing*, MCS shares many characteristics with it, but at the same time has several unique features. MCS typically involves *location dependency* (geo-tagged data) and *temporal continuity* (collecting data continuously over an extended period), and each individual worker only participates in a few *micro-tasks*. These features have significant impact on the incentive mechanism design, which we will elaborate. In particular, the temporal continuity also engenders the *sustainability* issue, where workers may not follow through the entire campaign but drop out in the interim. This is under-explored in the literature and is one of the foci of this article. A broader scope of sustainability, which encompasses other topics such as energy efficiency, security and privacy, warrants several other lines of rigorous research.

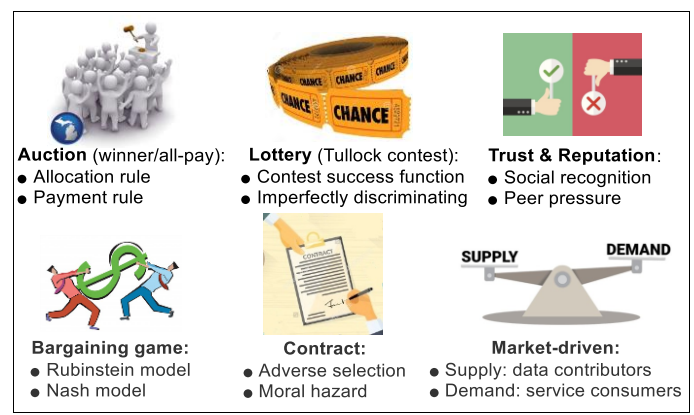

Fig. 1. Six incentive mechanism frameworks with their key elements.

## II. PRELIMINARIES OF MECHANISM DESIGN

Mechanism design concerns stipulating a set of rules such that players will act to the designer's preference. Therefore, mechanism design is also known as *reverse game theory*, since game theory concerns reasoning about players' strategy choices given a set of rules.

However, the space of designing the set of rules appears to be infinite, making the problem seemingly intractable. This issue was remarkably alleviated due to the introduction of the *revelation principle*, which says that any arbitrary mechanism can be replicated by an *incentive-compatible direct mechanism*. Here, a direct mechanism is one in which players directly tell their *types* (i.e., *private information* such as cost) to the designer, and is incentive-compatible (IC) if truth-telling is optimal for every player. Thus, the revelation principle allows us to restrict attention to direct mechanisms only, which are a much smaller class as compared to the original design space. Another important property that needs to be satisfied by a mechanism is *individual rationality* (IR), which means that one should only gain or maintain his/her utility by participating.

In practice, a player's type is often unknown to other agents and the mechanism designer, who hence have to reason about the unknowns using prior (often probabilistic) beliefs. This is called an *incomplete-information* setting and is dealt with by *Bayesian mechanism design*.

The classic mechanism design theory, which is rooted in economics, focuses on characterizing the existence and uniqueness of equilbria. Its recent marriage with computer science gave birth to the theory of *algorithmic mechanism design*, which focuses more on how to reach a desired equilibrium through polynomial-time algorithms, with an emphasis on *computational efficiency*.

### III. AUCTION

Auction is one of the most widely used incentive mechanism design frameworks in MCS. A standard auction consists of an auctioneer who sells some goods and a group of bidders who place bids to buy the goods. The auctioneer determines:
- An *allocation rule*, which specifies "who gets what," that is, who win the auction and what goods are allocated to them
- A *payment rule*, which dictates "who pays how much".

A classic example is a *Vickrey auction*, where there is a single good and the allocation rule is that the highest bidder gets the good, and the payment rule is that the highest bidder pays the second-highest bid. While seemingly simplistic, Vickrey auction possesses three very desirable properties: dominant-strategy incentive-compatibility, maximal social welfare, and computational efficiency.

When auctions are applied to MCS, the buyer and seller roles are often swapped: the bidders are now mobile users or *workers* who want to *sell* sensory data, and the auctioneer *buys* sensory data from them. This is often referred to as a *reverse auction* model, in which the allocation rule determines the winners (who are qualified to sell data) and the payment rule determines the size of payment to each winner.

Standard auctions can be categorized into *winner-pay* and *all-pay* auctions according to who pay the bids. To allow this taxonomy to also cover reverse auctions, we generalize "paying" bids to "fulfilling" bids. Thus, fulfilling a bid in standard auctions means paying one's bidding price, and in reverse auctions means surrendering a selling item. In MCS, the latter corresponds to completing a sensing task or submitting sensor data.

#### A. Winner-pay auctions

In a winner-pay auction, only the winners (selected by the allocation rule) need to pay or fulfill the bids. This conforms to intuition and has been widely applied in the MCS literature.

For example in [1], $n$ participants bid their respective desired payments $b_i$ for performing a sensing task requested by a service provider. The service provider (i.e., auctioneer) implements (a) an allocation rule that selects the lowest $m$ out of the $n$ bidders as the winners, and (b) a payment rule that pays the $m$ winners their respective bids $b_i$. This is essentially a *first-price sealed-bid auction* which does not satisfy IC, as a bidder could overbid ($b_i > c_i$ where $c_i$ is his true sensing cost) and gain higher payoff.

However, what is interesting in [1] is how the issue of *sustainability* is addressed. The authors observed that, as winner selection is based on $b_i$ which is lower bounded by $c_i$, the auction tends to separate participants into constant winners and losers according to their $c_i$ after multiple rounds. Thus, the loser group may start to *drop out* as they would see little chance to win. The resultant shrinking participant pool would then induce the winners to increase bids, which imply a higher cost to the service provider and impact the sustainability of the campaign.[1] To solve this problem, [1] gives "virtual participation credit" $\alpha$ to each losing participant after each round, such that his bid $b_i$ in the next round will be treated as $b_i' = b_i - k_i\alpha$ where $k_i$ is the number of his consecutive losing rounds. Hence, a loser gets higher chance to win subsequently while his payment remains $b_i$.

Another winner-pay auction that specifically addresses sustainability is [2], which selects winners by combining their locations (for better geographic coverage) and reported sensing costs. To provide long-term incentives, the auction aims to satisfy a *participatory constraint*: the average frequency that a user is selected must be no less than his "dropout threshold". Unlike [1], the auction [2] satisfies the IC constraint by adopting a *VCG auction*.

VCG auction is an extension of the classic Vickrey auction for selling multiple goods, which corresponds to allocating multiple sensing tasks in MCS. A VCG auction allocates goods to the set of bidders whose bids maximize the social welfare (total goods value); in MCS, this means allocating tasks to workers whose sensing costs minimize the total cost. As for the payment rule, each bidder $i$ pays his *externality*, i.e., the maximum welfare if $i$ were absent minus the current welfare (when $i$ is present) of others.

#### B. All-pay auctions

In an all-pay auction (APA), all the bidders need to pay or fulfill the bids regardless of who win the auction. This appears to be unnatural, and indeed, APA is rarely used in practice for selling traditional goods. But in fact, it exists in reality pervasively, but in a nonobvious form. For example, in political campaigns, job promotions, R&D competitions, and sports, all candidates exert vast effort (fulfilling bids) without knowing who will eventually win the competition. The theoretical foundation of APA is based on the notion of *expected utility*, which incorporates a *winning probability* into the utility function and thereby makes all-pay equivalent to winner-pay auctions in principle.

The first work that applies APA to MCS is [3], where the APA is conducted as follows. After a task requester announces a sensing task, interested workers can straightaway participate in performing the task (e.g., move to specific locations and collect sensor data). Upon completion of the task (or after a predefined period), the task requester selects a winner based on performance (amount/quality of collected sensor data) and rewards him. This is all-pay since non-winners have also surrendered their sensing data (and effort).

Compared to the winner-pay genre, APA has three desired advantages, as partially covered in [4]. The first is *simplicity*. A winner-pay auction consists of two stages (Fig. 2): a *bidding stage* in which bidders submit bids to indicate their *intent* to participate (e.g., how much sensor data to collect and how much payment they desire), and a *contribution stage* in which only the winners (a subset chosen from all the bidders) perform the sensing task. In contrast, an APA compresses these two stages into a single *bid-cum-contribution* stage, in which all workers contribute straightaway without bidding their intent. For a requester, such a MCS campaign is simpler to organize. For workers, they no longer need to contrive a plan or intent just for doing a micro-task like sensor data collection; rather, they can quickly start and then "plan on the go" (e.g., the amount of data). This offers more flexibility and is better suited to the ad-hoc and "micro-task" based nature of MCS.

---

[1] Strategic workers could do better by *underbidding* in earlier rounds so as to "elbow out" other workers and, thereafter, increase bids to gain higher payoff in the long run. However, in reality, workers are generally *myopic*, as also (implicitly) assumed by [1].

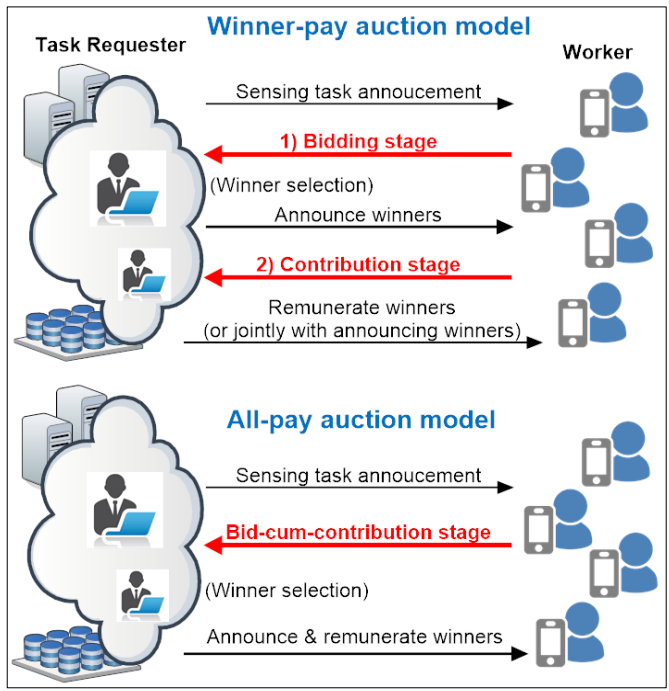

Fig. 2. Winner-pay auction vs. all-pay auction in MCS.

The second advantage is *risk-free* of bid nonfulfillment. Since a bid in winner-pay auctions is merely an intent to participate, there is little guarantee that the winning bids selected in the first stage will be fulfilled in the second stage. On the other hand, bids in APA are all fulfilled upfront (as actual contributions), which nullifies the risk.

The third advantage is *obliviousness to truthfulness*. This is a special merit when applying APA to MCS, as APA does not exhibit it by itself. This merit says that IC, which is a main challenge in mechanism design, is technically irrelevant to all-pay MCS. The reason is that both the allocation and payment rules of all-pay MCS are no longer based on bids of intent but on bids of actual contributions, which are directly observable and cannot be lied about. This liberates mechanism designers from the IC constraint and allows them to focus on other important goals such as revenue maximization, IR, and computational efficiency.

However, APA also has a disadvantage, which is more of a psychological rather than a technical one. That is, although the fact that APA entails a *sunk cost* to every bidder makes no mathematical difference in terms of *expected utility*, it demands the bidders to be fully *rational*. More specifically, APA can only guarantee nonnegative payoff *in expectation* but not on *every realization*, unlike winner-pay auctions. In other words, APA offers a weaker "sense of security" to workers. One remedy is to employ behavioral economics and marketing strategies, as suggested by [4].

Tab. 1 summarizes the above comparison.

TABLE 1. COMPARISON OF WINNER-PAY AND ALL-PAY MCS

|  | **Winner-pay MCS** | **All-pay MCS** |
|---|---|---|
| **Procedure** | Two stages: bidding and contribution | Single stage: bidding cum contribution |
| **Risk of bid nonfulfillment** | Yes | No |
| **Has challenge of satisfying IC** | Yes | No |
| **Workers' sense of security** | Stronger | Weaker |
| **Revenue (total contribution)** | Equal (by the *revenue equivalence theorem* under standard assumptions) ||

*C. Sustainability*

As mentioned in Section I, the temporal continuity of MCS causes a critical sustainability issue in which participants may drop out due to lack of long-term commitment. One way to retain participants is to run the original "grand" auction in multiple iterations, each over shorter periods, such that the remuneration cycle is reduced and more winners can be selected. However, under such a scheme, many workers may keep losing successive rounds and thus still quit in frustration.

Therefore, we suggest three modifications to traditional auction design to improve sustainability. First, redesign the *allocation rule* by determining winners using:
- The (possibly time-discounted) *cumulative contribution* of each non-winner rather than his contribution in the current round alone, or
- A *discriminatory winning probability*, which is a function of previous losing rounds, such that losers are "subsidized" with higher winning odds subsequently.

Second, redesign the *payment rule* such that the reward is *adaptive* to the losing history of a winner. An example can be found in [3][4], which introduce an *adaptive prize* to vary with winner's cumulative contribution so that workers are incentivized to contribute more than the case of fixed reward.

Third, although theory shows no definitive advantage between single and multiple prizes in terms of revenue (total contribution) [5], we recommend the use of multiple prizes for MCS. This is because it curbs "starvation", especially when the crowd size is large, and is user-friendly.

Another feature related to sustainability is the *microscopic nature of MCS tasks* as mentioned in Section I. This feature calls for a *minimal participation procedure*, as otherwise it tends to outweigh the task itself and thereby prompts participants to leave. This advocates all-pay auction as a more favorable choice due to its one-stage bidding process.

## IV. LOTTERY

Auctions have been extensively studied in economics for decades, and (primarily because of that) are widely adopted in the MCS literature as an incentive mechanism. However, a recent critique undertaken by [6] points out that auctions may not always be a good fit for MCS due to their *perfectly discriminating* nature. Intuitively, it means that one must outbid everyone else in order to win; in other words, auctions are so competitive that "weaker" (lower-type) bidders will *never* win. Thus, while auctions could be a superior choice for crowdsourcing that solicits prime quality from strong players, they may not be well suited to MCS which aims to engage "grassroots" to perform very simple tasks like sensor data collection, where *massive participation* is of the foremost priority to achieve a required geographic coverage.

*Lottery*---or its generalized form *Tullock contest*---is shown by [6] to be a good alternative to resolve this issue. A Tullock contest is a probabilistic game in which the winner is not determined by the rank of bids but by a probability, specified by a contest success function (CSF) $p_i = b_i^r / \sum_j b_j^r$. Here $b_i$ is bidder $i$'s bid and $r$ is a constant exponent. When $r=1$, it yields a lottery which is the simplest form of Tullock contests.

The most salient feature of Tullock contests is that they are *imperfectly discriminating*: as bids only determine winning probabilities, *everyone has a chance to win*, no matter how "weak" he is. This is very attractive to ordinary workers who often constitute the majority of MCS participants, which is not necessarily the case in crowdsourcing in general. Therefore, as evidenced even by reality, many countries run national lotteries in which millions of people participate.

Tab. 2 summarizes the above comparison, indicating that auction and lottery are two *complementary* mechanisms. When applied to MCS, a typical lottery is conducted in an all-pay fashion, in the sense that all the bids are actual contributions.

TABLE 2. COMPARISON OF AUCTION AND LOTTERY

|  | Auction | Lottery / Tullock Contest |
|---|---|---|
| **Winner selection** | Perfectly discriminating | Imperfectly discriminating |
| **Competitiveness / Barrier to entry** | High | Low |
| **Typical size of participant pool** | Small | Large |
| **Contribution level from each *individual* player** | High | Low |
| **Suitable applications** | Those favoring quality over quantity (e.g., effort/knowledge-intensive crowdsourcing, contests) | Those favoring quantity over quality (e.g., micro-task crowdsourcing, MCS) |
| **Suitable players** | Strong players (who are of higher types) | Ordinary players |
| **Revenue (total contribution)** | No conclusive comparison (contingent on problem settings) | |

Tullock contests are inherently more sustainable than auctions, because being imperfect discriminating allows for a more even distribution of winning positions and thereby helps participant retention. To further improve sustainability, one way is to incorporate historical losing records into the bid $b_i$ or the power exponent $r$ in the CSF, such that the CSF gives favorable bias toward continual losers. Another way is to use the *adaptive payment rule* described in Section III, for which [6] provides a detailed reference.

## V. TRUST AND REPUTATION SYSTEMS

Auctions and Tullock contests tend to use financial incentives, which may be less effective when:
- The amount is insignificant (for example due to the "micro-ness" of MCS tasks), or
- The task has moral implications (e.g., collecting healthcare-related data for seniors).

Another issue is *financial sustainability*, which we address in Section VI-C.

A widely used non-monetary incentive mechanism is trust and reputation systems. Trust is a local and subjective measure of relationship between two persons/agents, and can be derived from direct or indirect past interactions. Reputation is a global and rather objective measure by aggregating all other people's trust with respect to a particular person. Trust and reputation have enormous influence on *social recognition* and *peer pressure*, and hence are effective and sustaining sources of motivation as backed up by both scientific research and practice (e.g., Quora and StackOverflow).

A well-known online trust and reputation system is the *Beta reputation system* [7]. It uses a modified expected value of the Beta distribution to model the extent to which a user $i$ trusts another user $j$, as $T(i,j)=[g(i,j)-b(i,j)]/[g(i,j)+b(i,j)+2]$, where $g(i,j)$ and $b(i,j)$ are the number of "good" and "bad" feedbacks $i$ gave to $j$, respectively. The reputation of $j$ is then an aggregated value of all the feedback combined, i.e., $R_j = (g_j-b_j)/(g_j+b_j+2)$ where $g_j=\sum_i g(i,j)$ and $b_j=\sum_i b(i,j)$.

Trust and reputation can be used in MCS to incentivize workers to contribute more trustworthy data. For example in [8], the authors use a fuzzy inference system to determine the trust-of-contribution, given the quality of contributed data and the trust-of-participant. If the output trust is higher than a threshold, the reputation of the participant will increase, otherwise it will decrease. The reputation is then used as a scaling factor of reward, thereby incentivizing each worker to improve his quality of contribution and to contribute more.

Another trust and reputation based incentive mechanism for MCS is *simple endorsement web* (SEW) [9]. It is a *social network* that connects participants using an *endorsement* relationship, where Alice endorses Bob if she trusts Bob to be a "good" contributor, or because of benefit derived from *nepotism*.

Nepotism is a notion introduced by [9] to capture human nature more realistically, striking a tradeoff between *egoism* (as assumed by game-theoretical economists) and *altruism* (as argued for by philanthropists and humanitarians). Nepotism states that people could behave in the interest of a specific group of people whom they care about (e.g., family and close social connections), rather than being categorically egoistic or altruistic (Fig. 3).

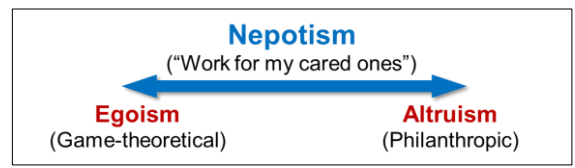

Fig. 3. *Nepotism* [9] strikes a tradeoff between egoism and altruism, aiming to capture human nature in a more realistic manner.

Nepotism can be used in social-network based MCS to create incentives. For example, this can be realized by a *revenue-sharing scheme* [9] in which a worker who contributes sensor data and thereby earns reward (e.g., reputation points) needs to share the reward with his endorsers. As a result, endorsers become *beneficiaries* of the contributor, and if they include his "nepotic" social connections, a new incentive is created for the contributor: "*work for your cared (or loved) ones*" (besides yourself).

For completeness (to cover non-nepotic cases as well), endorsement is designed to be a *mutual-beneficial* relationship, where a contributor with more endorsers is deemed by SEW as more trustworthy, and will receive higher reward.

Trust and reputation systems are generally more sustainable than monetary incentive mechanisms, due to the void of financial burden and the long-term social influence. On the other hand, a major challenge to trust and reputation systems is the *cold-start problem*, i.e., the difficulty of inferring the trustworthiness of a user during bootstrapping. There is a large literature on this topic which is out of the scope of this paper.

## VI. OTHER INCENTIVE MECHANISMS

We discuss three additional incentive mechanisms which are less common in the MCS literature but have great potential nevertheless.

### A. Bargaining games

A bargaining game concerns how to divide certain surplus (cooperation benefit) between two players. There are two classic bargaining models. The *Rubinstein bargaining model* takes a strategic approach to model the bargaining procedure as a *sequential game*, in which the two players alternately propose offers until one accepts the offer proposed by the other. The *Nash bargaining model* takes an axiomatic approach to focus on deriving an outcome that satisfies certain axioms [10]. Such an outcome is a tuple ($r_1$, $r_2$) that maximizes the *Nash product*, $(u_1(r_1)-u_1(d_1))(u_2(r_2)-u_2(d_2))$, where respectively, $r_1$ and $r_2$ are the two players' shares of the total surplus, $u_1(.)$ and $u_2(.)$ are their utility functions, and $d_1$ and $d_2$ are their status quo payoffs if an agreement is *not* achieved.

MCS involves multiple workers, and we can apply a bargaining model by letting the task requester bargain with each worker separately while taking into account other workers. Using the Nash model as an example (e.g., [11]), let us suppose a task requester has a sensing task of value $v$ and a worker is interested to undertake it. The requester wants a share $r_1$ as profit and the worker wants a share $r_2$ as reward, where $r_1+r_2 \leq v$. If the bargain fails (say with a probability $p$), they both fall back to their status quo payoffs, which are typically 0 for the worker but can be positive for the requester. The reason is that, with probability $1-p$, the requester can reach an agreement with one of the other workers. This implies that $d_1>0$ and $d_2=0$, which then allows us to formulate and optimize the Nash product. Depending on the players' risk profiles, the utility functions $u_1(.)$ and $u_2(.)$ may be nonlinear.

The sustainability issue arises when some workers constantly fail to achieve an agreement with the requester and fall back to their status quo payoffs. This can be improved by giving higher *bargaining power* to "loyal" workers and workers who successively fail. One way to achieve this is to generalize the Nash product to $(u_1(r_1)-u_1(d_1))^{\alpha}(u_2(r_2)-u_2(d_2))^{1-\alpha}$ to create the *asymmetric bargaining power* case, where $\alpha<0.5$ gives an advantage to player 2. Another way is to make the worker's status quo payoff $d_2$ a function of his/her loyalty or bargaining history.

Moreover, as the bargaining process itself creates no surplus but can be costly, we can improve sustainability by *automating* the bargaining process using software. This is a non-theoretical tweak but can be very useful in practice.

### B. Contract theory

Contract theory [12] deals with two players who take very different roles. One player, called a *principal*, has all the bargaining power and spells out a contract, which may contain a list of contract items. The other, called an *agent*, can only accept or reject the contract or accept a specific contract item, without counter-offering like in bargaining.

There are two main contract models: the *adverse selection* model, in which the agent has certain *hidden information* that the principal tries to elicit, and the *moral hazard* model, in which the agent could exert some *hidden effort* that is of economic value to the principal, and the principal tries to induce a desired effort level at minimal cost.

In the context of MCS, we consider each pairing of a worker and the task requester. In the adverse selection case, the hidden information may be the worker's sensing cost. The requester can offer a menu of contract items each being a (cost, remuneration) tuple. To induce the worker to pick the contract item corresponding to his/her true sensing cost (i.e., to satisfy IC), the requester needs to pay an *information rent* which is the difference between the remuneration and the cost. See [13] for an example of how to set the rent. In the moral hazard model, the requester aims to elicit certain sensing effort from a worker so as to produce a sensing quality that maximizes the data value $v$ minus the worker's remuneration. As the effort is hidden and the quality is not a deterministic function of effort, the optimal contract is an *insurance* that gives the worker the entire, effort-dependent $v$ while requiring the worker to pay a fixed "deductible". However, if the worker is risk-averse, the requester needs to also offer a quality-linked incentive and make a tradeoff between incentive and insurance based on the worker's risk profile.

Sustainable contracts can be achieved in two ways. First, the contract can adopt an *installment scheme* rather than one-off payment: only after the worker has collected a certain portion of the total target amount of sensing data (with certain quality), will a corresponding portion of the total remuneration be paid to the worker. This not only motivates workers to follow through the entire campaign, but also shortens their waiting period and curbs impatience. Second, in the adverse selection model, the remuneration can include a bonus component on top of information rent to reward long-term workers; in the moral hazard model, this bonus can be incorporated into either the insurance or the incentive.

### C. Market-drivern mechanisms

Monetary incentives may encounter *financial sustainability* as mentioned in Section V, where constant payments to workers could impose stringent burden on budget. One solution is *market-driven mechanisms*, which exploit the *supply-demand* interaction to create incentives and shed financial burden from MCS systems.

To run a market-driven mechanism, the MCS system first needs to create a market. Specifically, given that supply is provided by MCS workers, the goal is to create *demand*, i.e., attract *consumers*. This can be achieved by (i) offering a compelling informational service over the collected sensor data (e.g., via data analytics), or (ii) simply providing the raw data if it bears considerable value to certain users.

The next step is to design a market-driven mechanism using one of the following models. In a *fine-grained* model, each service request from a consumer can be mapped to a specific set of data contributions. For example, a consumer may query "the average traffic speed of Road-7 in the past hour". In such cases, the market can distribute the consumer's payment to workers who contributed data to that particular spatio-temporal (S-T) window. As illustrated in Fig. 4, a requester who made a query at S-T point $t_1^s$ pays workers who made the set of contributions $\{q_{(1)}, q_{(2)}, q_{(3)}\}$, and a query at $t_2^s$ pays to the set $\{q_{(3)}, q_{(4)}\}$. *Dynamic pricing* [14] can also be integrated to determine the payment to each individual worker.

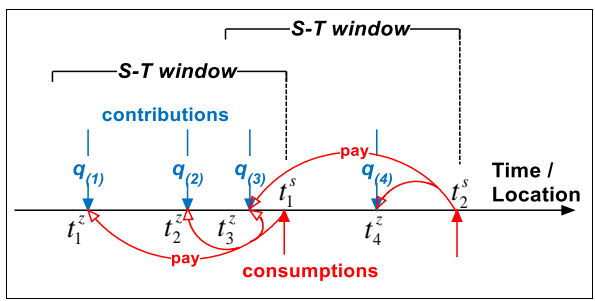

Fig. 4. Fine-grained model for market-driven mechanisms [14].

In a *coarse-grained* model, each consumption is serviced by mining a large set of data or multi-datasets; it is not possible or practical to pinpoint which particular contributions are used and to what extent. Therefore, supply and demand can be calculated on a macro basis using statistical methods to determine pricing and payment distribution.

In both models, it is possible that each user plays a dual role of both data contributors and service consumers. A corresponding incentive scheme is provided in [15], which does not use monetary payments.

Market-driven mechanisms thus improve sustainability by shedding financial burden from the system. They are also advantageous in their inherent ability to regulate *supply imbalance* between popular and unpopular areas, or peak and non-peak hours. This is achieved by charging a higher price to spatiotemporal regions with lower supply but higher demand, and vice versa, which incentivizes workers to move to system-desired regions to perform MCS tasks.

VII. SUMMARY, CHALLENGES AND OPPORTUNITIES

Is there a rule of thumb as to which incentive mechanism fits which particular MCS applications? The answer is embedded above and summarized here. In general, auctions suit effort/knowledge-intensive applications while lotteries suit micro-task scenarios. Trust and reputation systems are best when the task has strong moral and social implications, while market-driven mechanisms are a superior choice when the sensing data have great commercial value; both mechanisms have good financial sustainability. Bargaining games suit the situation when workers and the task requester have comparable bargaining power, while contracts are preferred when the task requester dominates the decision making.

While research in the area of incentive mechanism design is rich, the fundamental assumption of human rationality often faces challenge in reality. A relaxation of this assumption is the notion of *bounded rationality*, which has led to rising activities on *behavioral economics*. Another challenge is *collusion* among agents, which significantly complicates the design but meanwhile introduces a very interesting problem to solve. *Heterogeneity* and *inter-correlation* of agent types pose additional challenges by often precluding closed-form solutions. Moreover, non-quasilinear utility functions and uncertain risk-profiles have been much less studied and are worth future exploration.